\documentclass[conference]{IEEEtran}

\usepackage{cite}
\usepackage{amsmath}
\usepackage{pifont}
\usepackage{textcomp}
\usepackage{mathtools}
\usepackage{algorithmic}
\usepackage{varwidth}
\usepackage{tabularx}
\usepackage{listings}

\usepackage{array}
\ifCLASSOPTIONcompsoc
  \usepackage[caption=false,font=normalsize,labelfont=sf,textfont=sf]{subfig}
\else
  \usepackage[caption=false,font=footnotesize]{subfig}
\fi
\usepackage{stfloats}
\usepackage{url}
\usepackage[colorlinks=true,
            linkcolor=blue,
            urlcolor=blue,
            citecolor=blue]{hyperref}

\usepackage{orcidlink}

\usepackage[utf8]{inputenc}
\DeclareUnicodeCharacter{1EC5}{$\tilde{\hat{\text{e}}}$} % U+1EC5 = ễ

\usepackage[normalem]{ulem}

\usepackage{xcolor}

\definecolor{codebg}{rgb}{0.98,0.98,0.98}
\definecolor{codekw}{HTML}{0000BB} % blue
\definecolor{codestr}{HTML}{A31515} % red-ish
\definecolor{codecm}{HTML}{008000} % green
\definecolor{codenum}{HTML}{098658} % green-ish

\definecolor{code1}{rgb}{0.7,0.9,0.9}
\definecolor{code2}{rgb}{0.9,0.9,0.7}
\definecolor{code3}{rgb}{0.9,0.7,0.9}

\definecolor{codemove}{rgb}{0.8,0.8,0.9}
\definecolor{codeplus}{rgb}{0.8,0.9,0.8}
\definecolor{codeplusplus}{rgb}{0.6,0.9,0.6}
\definecolor{codeminus}{rgb}{0.9,0.8,0.8}
\definecolor{codeminusminus}{rgb}{0.9,0.6,0.6}

\definecolor{codeA}{rgb}{0.7,0.9,0.9}
\definecolor{codeB}{rgb}{0.9,0.9,0.7}
\definecolor{codeC}{rgb}{0.9,0.7,0.9}
\definecolor{codeD}{rgb}{0.7,0.7,0.7}
\definecolor{codeE}{rgb}{0.9,0.7,0.7}
\definecolor{codeF}{rgb}{0.7,0.9,0.7}
\definecolor{codeG}{rgb}{0.5,0.7,0.9}
\definecolor{codeH}{rgb}{0.5,0.9,0.7}
\definecolor{codeI}{rgb}{0.7,0.5,0.9}
\definecolor{codeJ}{rgb}{0.7,0.9,0.7}
\definecolor{codeK}{rgb}{0.9,0.5,0.7}
\definecolor{codeL}{rgb}{0.9,0.7,0.5}

\definecolor{codeM}{rgb}{1.0,1.0,1.0}
\definecolor{codeN}{rgb}{1.0,1.0,1.0}
\definecolor{codeO}{rgb}{1.0,1.0,1.0}
\definecolor{codeP}{rgb}{1.0,1.0,1.0}
\definecolor{codeQ}{rgb}{1.0,1.0,1.0}
\definecolor{codeR}{rgb}{1.0,1.0,1.0}
\definecolor{codeS}{rgb}{1.0,1.0,1.0}
\definecolor{codeT}{rgb}{1.0,1.0,1.0}
\definecolor{codeU}{rgb}{1.0,1.0,1.0}
\definecolor{codeV}{rgb}{1.0,1.0,1.0}
\definecolor{codeW}{rgb}{1.0,1.0,1.0}
\definecolor{codeX}{rgb}{1.0,1.0,1.0}
\definecolor{codeY}{rgb}{1.0,1.0,1.0}
\definecolor{codeZ}{rgb}{1.0,1.0,1.0}

\newcommand{\codebg}[0]{%
  \color{codebg}%
}
\newcommand{\codekw}[0]{%
  \color{codekw}\bfseries%
}
\newcommand{\codecm}[0]{%
  \color{codecm}\itshape%
}
\newcommand{\codestr}[0]{%
  \color{codestr}%
}

\lstdefinestyle{ide}{
  backgroundcolor=\codebg,
  keywordstyle=\codekw,
  commentstyle=\codecm,
  stringstyle=\codestr,
  stepnumber=1,
  showstringspaces=false,
  breaklines=true,
  tabsize=5,
}

\usepackage{tikz}
\usetikzlibrary{calc}

\renewcommand{\dashbox}[2][black]{%
  \tikz[baseline=(X.base)]%
    \node[
      draw=#1,          % border color
      dash pattern=on 2.5pt off 1pt,           % dashed border
      inner sep=1pt,    % padding
      rounded corners=2pt
    ] (X) {#2};%
}

\newsavebox{\codehlboxbox}
\newcommand{\codehl}[2]{%
  {%
    \sbox{\codehlboxbox}{\begin{tabular}{@{}l@{}}#2\end{tabular}}%
    \setlength{\fboxsep}{0pt}%
    \colorbox{#1}{\usebox{\codehlboxbox}}%
  }%
}

\newcommand*{\op}[1]{\codehl{code#1}{$\mathcal{#1}$}}

\newcommand{\targetbox}[2]{%
  {%
    \sbox{\codehlboxbox}{\begin{tabular}{@{}l@{}}#2\end{tabular}}%
    \setlength{\fboxsep}{1pt}%
    \dashbox{\usebox{\codehlboxbox}}\dashbox{$\op{#1}$}%
  }%
}

\newcommand{\codemove}[1]{%
  \codehl{codemove}{#1}%
}

\newcommand{\codeplus}[1]{%
  \codehl{codeplus}{#1}%
}

\newcommand{\codeminus}[1]{%
  \codehl{codeminus}{#1}%
}

\newcommand{\codeplusplus}[1]{%
  \codehl{codeplusplus}{#1}%
}

\newcommand{\codeminusminus}[1]{%
  \codehl{codeminusminus}{#1}%
}

\newcommand{\codeminusminusstrike}[1]{%
  \codehl{codeminusminus}{\sout{#1}}%
}

\lstdefinestyle{mycode}{
  language=Python,
  basicstyle=\tiny\ttfamily, % Global font style (monospaced, small)
  style=ide,
  numberstyle=\tiny\color{gray}, % Style of the line numbers
  breaklines=true, % Allow long lines to break automatically
  columns=fullflexible,
}

\usepackage{booktabs}

\newenvironment{mycode}{%
  \providecommand{\tab}[0]{\phantom{~~~~}}%
  \providecommand{\kw}[1]{{\codekw{##1}}}%
  \providecommand{\mov}[1]{{\codemove{##1}}}%
  \providecommand{\add}[1]{{\codeplus{##1}}}%
  \providecommand{\ADD}[1]{{\codeplusplus{##1}}}%
  \providecommand{\rem}[1]{{\codeminus{##1}}}%
  \providecommand{\REM}[1]{{\codeminusminus{##1}}}%
  \providecommand{\REMST}[1]{{\codeminusminusstrike{##1}}}%
  \providecommand{\tgt}[2]{{\targetbox{##1}{##2}}}%
  \noindent\tiny\ttfamily%
  \begin{tabular}{@{}l@{}}%
}{%
  \end{tabular}%
}

\usepackage{xspace}

\usepackage[english]{babel}
\usepackage{csquotes}

\MakeOuterQuote{"}

\newcommand{\betweentinyandscriptsize}{\fontsize{6pt}{7.2pt}\selectfont}

\let\origtexttt\texttt
\newcommand{\textttsize}{\footnotesize}
\renewcommand{\texttt}[1]{{\textttsize\origtexttt{#1}}}

\newcommand{\scastt}[1]{
  \text{\ttfamily\betweentinyandscriptsize #1}
}

\newcommand{\scasarrowmath}[1]{
  \xrightarrow{\scriptsize \text{}#1\text{}}
}

\newcommand{\scasarrow}[1]{
  $\scasarrowmath{#1}$
}

\newcommand{\scasdownarrowmath}[2][]{%
  \downarrow #1 \text{\scriptsize $#2$}%
}

\newcommand{\scasdownarrowaftermath}[2][]{%
  \text{\scriptsize $#2$} \downarrow #1%
}

\newcommand{\scasdownarrowinner}[2][]{%
  $\scasdownarrowmath[#1]{#2}$%
}

\newcommand{\scasdownarrowafterinner}[2][]{%
  $\scasdownarrowaftermath[#1]{#2}$%
}

\newcommand{\opprop}[1]{\textsc{#1}\xspace}

\newcommand{\Complete}{\opprop{Complete}}
\newcommand{\Composable}{\opprop{Composable}}
\newcommand{\Deterministic}{\opprop{Deterministic}}

\newcommand{\Nullipotent}{\opprop{Nullipotent}}
\newcommand{\Pluripotent}{\opprop{Pluripotent}}
\newcommand{\Commutative}{\opprop{Commutative}}
\newcommand{\Commutativity}{\opprop{Commutativity}}

\newcommand{\opname}[1]{\ensuremath{\mathit{#1}}\xspace}

\newcommand{\AddParam}{\opname{AddParam}}
\newcommand{\MakeCond}{\opname{MakeCond}}
\newcommand{\MakeOptional}{\opname{MakeOptional}}
\newcommand{\Rename}{\opname{Rename}}
\newcommand{\GetEnclosingFn}{\opname{GetEnclosingFn}}
\newcommand{\MoveParam}{\opname{MoveParam}}
\newcommand{\TupleToStruct}{\opname{TupleToStruct}}
\newcommand{\WrapFunction}{\opname{WrapFunction}}

\begin{document}
\bstctlcite{IEEE:BSTcontrol}

%%%%%%%%%%%%%%%%%%%%%%%%%%%%%%%%%
%%
%%  TITLE AND AUTHOR
%%
%%%%%%%%%%%%%%%%%%%%%%%%%%%%%%%%%

% paper title
% Titles are generally capitalized except for words such as a, an, and, as,
% at, but, by, for, in, nor, of, on, or, the, to and up, which are usually
% not capitalized unless they are the first or last word of the title.
% Linebreaks \\ can be used within to get better formatting as desired.
% Do not put math or special symbols in the title.
\title{Beyond Text Editing:\\Algebraic Manipulation of Source Code}

% author names and affiliations
% use a multiple column layout for up to three different
% affiliations
\author{\IEEEauthorblockN{Kevin Pulo\orcidlink{0009-0006-5903-6376}}
\IEEEauthorblockA{%
MongoDB Research%
\\
\href{mailto:kevin.pulo@mongodb.com}{kevin.pulo@mongodb.com}%
}}

% make the title area
\maketitle

%%%%%%%%%%%%%%%%%%%%%%%%%%%%%%%%%
%%
%%  ABSTRACT
%%
%%%%%%%%%%%%%%%%%%%%%%%%%%%%%%%%%

% As a general rule, do not put math, special symbols or citations
% in the abstract
\begin{abstract}
Source code is almost universally edited as plain text.  However, the mismatch between the syntactic and semantic requirements of valid and correct code, and the unconstrained text editing process trying to produce it, introduces friction that degrades the programming task.  It is also increasingly costly in the era of LLM-based coding agents, which must materialize their high-level plan of intended changes as low-level text edits dispersed throughout the codebase, often requiring them to re-read large portions of code.

We propose the novel alternative approach of \emph{source code algebra}, where the codebase is modified by applying to it a sequence of \emph{logical algebraic operations}.  Each operation makes the full set of changes necessary for a single semantic change, analogous to \emph{mathematical equation rewriting}.  We sketch initial properties of such operations --- including composition, nullipotency, and commutativity --- that distinguish this approach from text editing, and make it well-suited as a substrate for agentic code editing.

A feasibility probe with our proof-of-concept implementation (SCAS) suggests that LLM agents can use source code algebra to complete a non-local, cross-file code change with both higher success rates and one to two orders of magnitude fewer tokens, relative to text-based baselines.  While preliminary, this is consistent with the hypothesis that having LLMs emit algebraic operations, rather than rewritten code, is a promising direction for code editing --- and motivates broader future research into source code algebra, such as comprehensive operator libraries, formal properties, and human-facing tooling.
\end{abstract}

% Keywords:
%   Source code algebra
%   LLM-based code editing
%   Structural code editing
%   Source code manipulation
%   Software refactoring
%   Software evolution
%   Programming

% Topics:
%   Large Language Models for software evolution and maintenance tasks
%   Software refactoring and restructuring
%   Source code analysis and manipulation

% For peerreview papers, this IEEEtran command inserts a page break and
% creates the second title. It will be ignored for other modes.
\IEEEpeerreviewmaketitle

%%%%%%%%%%%%%%%%%%%%%%%%%%%%%%%%%
%%
%%  INTRODUCTION
%%
%%%%%%%%%%%%%%%%%%%%%%%%%%%%%%%%%

\section{Introduction}
\label{sec:intro}

\begingroup

\newsavebox{\boxcodea}
\begin{lrbox}{\boxcodea}
\begin{mycode}%
\kw{def} \tgt{A}{score(hits, misses)}:\\
\tab total = hits + misses\\
\tab accuracy = hits / total\\
\tab bonus = 0\\
\tab \kw{if} accuracy > 0.9: bonus = 10\\
\tab \kw{elif} accuracy > 0.8: bonus = 5\\
\tab \kw{return} hits * 2 - misses + bonus\\
\kw{print}(score(10, 3))\\
\kw{print}(score(23, 1))
\end{mycode}
\end{lrbox}

\newsavebox{\boxlabela}
\begin{lrbox}{\boxlabela}
\scriptsize
\begin{tabular}{@{}l@{}}
$\AddParam(\op{A},$~~~\\
$~~~~~~\scastt{includeBonus},$\\
$~~~~~~\scastt{True})$\\
\end{tabular}
\end{lrbox}

\newsavebox{\boxcodeb}
\begin{lrbox}{\boxcodeb}
\begin{mycode}%
\kw{def} score(hits, misses\add{, includeBonus}):\\
\tab total = hits + misses\\
\tab accuracy = hits / total\\
\tab bonus = 0\\
\tgt{B}{%
\tab \kw{if} accuracy > 0.9: bonus = 10\\
\tab \kw{elif} accuracy > 0.8: bonus = 5\\
}\\
\tab \kw{return} hits * 2 - misses + bonus\\
\kw{print}(score(10, 3\add{, \kw{True}}))\\
\kw{print}(score(23, 1\add{, \kw{True}}))
\end{mycode}
\end{lrbox}

\newsavebox{\boxlabelb}
\begin{lrbox}{\boxlabelb}
\scriptsize
\begin{tabular}{@{}l@{}}
$\MakeCond(\op{B},$~~~\\
$~~~~~~\scastt{includeBonus})$\\
\end{tabular}
\end{lrbox}

\newsavebox{\boxcodec}
\begin{lrbox}{\boxcodec}
\begin{mycode}%
\kw{def} score(hits, misses, includeBonus):\\
\tab total = hits + misses\\
\tab accuracy = hits / total\\
\tab bonus = 0\\
\tab \add{\kw{if} includeBonus:}\\
\tab \add{\tab}\mov{\kw{if} accuracy > 0.9: bonus = 10}\\
\tab \add{\tab}\mov{\kw{elif} accuracy > 0.8: bonus = 5}\\
\tab \kw{return} hits * 2 - misses + bonus\\
\kw{print}(score(10, 3, \kw{True}))\\
\kw{print}(score(23, 1, \kw{True}))
\end{mycode}
\end{lrbox}

\newsavebox{\boxcoded}
\begin{lrbox}{\boxcoded}
\begin{mycode}%
\kw{def} score(hits, misses):\\
\tab total = hits + misses\\
\tab accuracy = hits / total\\
\tab bonus = 0\\
\tgt{D}{%
\tab \kw{if} accuracy > 0.9: bonus = 10\\
\tab \kw{elif} accuracy > 0.8: bonus = 5\\
}\\
\tab \kw{return} hits * 2 - misses + bonus\\
\kw{print}(score(10, 3))\\
\kw{print}(score(23, 1))
\end{mycode}
\end{lrbox}

\newsavebox{\compositioncodebox}
\begin{lrbox}{\compositioncodebox}
\betweentinyandscriptsize
\begin{varwidth}{\linewidth}
\begin{algorithmic}
  \STATE \textbf{procedure} \MakeOptional(\op{T}, name)
  \STATE \quad \quad \AddParam(\GetEnclosingFn(\op{T}), name, \textbf{true})
  \STATE \quad \quad \MakeCond(\op{T}, name)
  \STATE \textbf{end procedure}
  \STATE
\end{algorithmic}
\end{varwidth}
\end{lrbox}

\newsavebox{\compositionbox}
\begin{lrbox}{\compositionbox}
\begin{tabular}{@{}l@{}}
\raisebox{-3ex}{\usebox{\compositioncodebox}}%
\\
\scasarrow{%
~~~~~~~~~~~~~~~~~~~~~~~
\MakeOptional(\op{D}, \scastt{includeBonus})%
~~~~~~~~~~~~~~~~~~~~~~~
}%
\end{tabular}%
\end{lrbox}

\newsavebox{\boxcodee}
\begin{lrbox}{\boxcodee}
\begin{mycode}%
\kw{def} score(hits, misses\add{, includeBonus}):\\
\tab total = hits + misses\\
\tab accuracy = hits / total\\
\tab bonus = 0\\
\tab \add{\kw{if} includeBonus:}\\
\tab \add{\tab}\mov{\kw{if} accuracy > 0.9: bonus = 10}\\
\tab \add{\tab}\mov{\kw{elif} accuracy > 0.8: bonus = 5}\\
\tab \kw{return} hits * 2 - misses + bonus\\
\kw{print}(score(10, 3\add{, \kw{True}}))\\
\kw{print}(score(23, 1\add{, \kw{True}}))
\end{mycode}
\end{lrbox}

\begin{figure*}[!b]
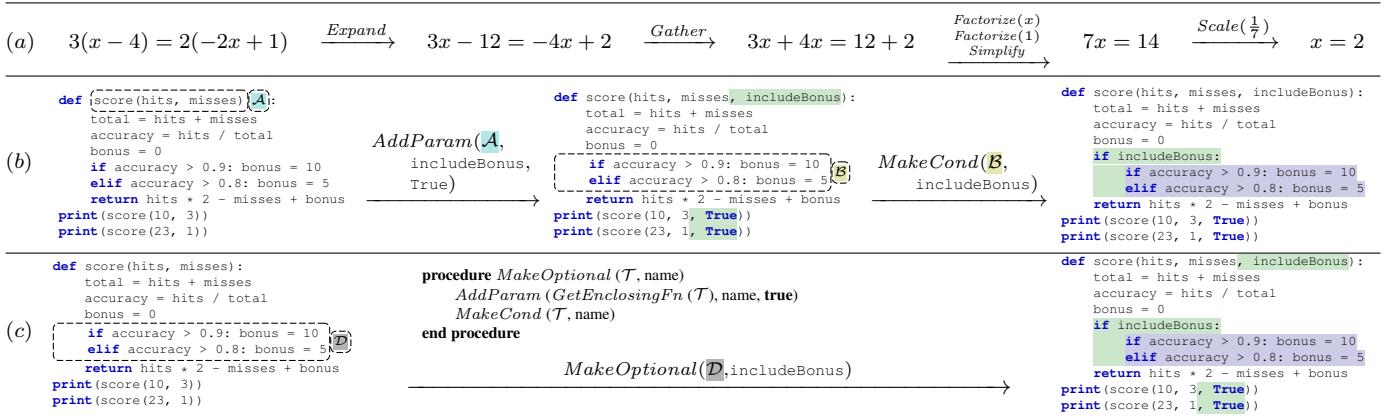

{
\footnotesize
\begin{tabular}{@{}l@{\;\;\;}c@{\;}c@{\;}c@{\;}c@{\;}c}
\hline %%%%%%%%%%%%%%%%%%%%%%%%%%%%%%%%%%%%%%%%%%%%%%%%%%%%%%%%%%%%%%%%%%%%%%%%%%%%%%
\noalign{\smallskip}

$(a)$
&
\multicolumn{5}{c}{
$
3(x - 4) = 2(-2x + 1)
\hfill
\scasarrowmath{\opname{Expand}}
\hfill
3x - 12 = -4x + 2
\hfill
\scasarrowmath{\opname{Gather}}
\hfill
3x + 4x = 12 + 2
\hfill
\raisebox{-2ex}{%
\scasarrow{%
\substack{%
\opname{Factorize}(x) \\%
\opname{Factorize}(1) \\%
\opname{Simplify}%
}%
}%
}
\hfill
7x = 14
\hfill
\scasarrowmath{\opname{Scale}(\frac{1}{7})}
\hfill
x = 2
$
}
\\%
\noalign{\smallskip}
\hline %%%%%%%%%%%%%%%%%%%%%%%%%%%%%%%%%%%%%%%%%%%%%%%%%%%%%%%%%%%%%%%%%%%%%%%%%%%%%%
\noalign{\smallskip}

$(b)$
&
\usebox{\boxcodea}
&
\raisebox{-4ex}{\scasarrow{\usebox{\boxlabela}}}
&
\usebox{\boxcodeb}
&
\raisebox{-4ex}{\scasarrow{\usebox{\boxlabelb}}}
&
\usebox{\boxcodec}

\\
\noalign{\smallskip}
\hline %%%%%%%%%%%%%%%%%%%%%%%%%%%%%%%%%%%%%%%%%%%%%%%%%%%%%%%%%%%%%%%%%%%%%%%%%%%%%%

$(c)$
&
\usebox{\boxcoded}
&
\multicolumn{3}{c}
{
\usebox{\compositionbox}%
}%
&
\usebox{\boxcodee}

\\
\noalign{\smallskip}
\hline %%%%%%%%%%%%%%%%%%%%%%%%%%%%%%%%%%%%%%%%%%%%%%%%%%%%%%%%%%%%%%%%%%%%%%%%%%%%%%

\end{tabular}
}

\caption{%
$(a)$~Mathematical equations are manipulated by iteratively rewriting them using algebraic identities and properties, leading to a rigorous and methodical process.
\quad
$(b)$~\emph{Source code algebra} extends this idea by treating the entire codebase as "the equation", allowing the programmer to manipulate the source code by applying global logical transformation operations to it.  Calligraphic parameters to operations (e.g. \op{A}) denote the targeted region of code.  Output colors indicate \codehl{codeplus}{added} or \codehl{codemove}{moved} code.  Here, the \AddParam operation adds a new parameter to a function, including updating all calls to the function, and the \MakeCond operation makes the targeted code execute conditionally.
\quad
$(c)$~Like mathematical algebra, operations can be composed to create more sophisticated source code manipulations.  This allows the programmer to operate closer to their higher-level mental plan of intended code changes.  Here, a new operation \MakeOptional is defined, which abstracts the changes from $(b)$ into the higher-level concept of "make the targeted code execute optionally, based on a new parameter of the enclosing function".
}
\label{fig:intro}
\end{figure*}

\endgroup

LLM-based coding agents are now a routine part of software development, but the way they modify source code has not fundamentally changed: agents read text, plan a change, and emit text edits --- sometimes via patches, sometimes by rewriting whole files, sometimes guided by sophisticated indexing and retrieval.  The agent must identify every location that needs to change, materialize each edit in a syntactically and semantically valid way, and keep enough of the codebase in context to do this reliably, even though that can mean re-reading large portions of code repeatedly.

This burden is not new.  Human programmers have always faced it, despite advances such as IDEs, syntax highlighting, code completion, static analysis, and refactoring tools.  The unconstrained nature of text editing can lead to deficiencies in the resulting code: not just typos and syntax errors (which are often easily found and fixed), but more significant problems from the mental effort required to text-edit at scale --- failing to cascade the downstream implications of changes, or subtle logic errors from focus being scattered over multiple in-flight changes, or across many locations in the codebase.

This paper presents a novel alternative.  Rather than modifying source code by editing its text representation, the programmer (human or LLM) interactively applies a sequence of \emph{logical "source code algebra" operations} to the entire codebase.  Each operation makes all the changes necessary for a single semantic step, and a sequence of operations can produce any new feature, bugfix, optimization, refactoring, or other desired change.  For example, an individual operation might convert a tuple to a struct (\TupleToStruct), introduce a wrapper around a function (\WrapFunction), or make some statements execute conditionally (\MakeCond).  A source code algebra system provides a library of such operations.

This approach is inspired by mathematical equation rewriting, where the mathematician applies strategically-chosen operations that take advantage of algebraic properties to carefully and gradually transform the equation into the desired form.  Each modification is small enough to verify, allowing their composition to be shown correct.

The algebraic approach to programming is analogous — the entire codebase is the "equation", and instead of tinkering with it ad-hoc in a text editor, the programmer systematically applies operations, each of which provides certain guarantees about the resulting code.

Fig.~\ref{fig:intro} illustrates this concept.  Fig.~\ref{fig:intro}(a) shows the familiar arithmetic algebra concept to find a solution for the value of $x$ in the equation.  Fig.~\ref{fig:intro}(b) presents an example Python function for performing a ``score'' calculation.  This source code is modified in two stages.

First, a new parameter is added to the \texttt{score} function called \texttt{includeBonus}.  This is achieved by applying the \AddParam operation to the function signature declaration \op{A}, which adds the parameter to the function signature, but is also responsible for adjusting all call sites to \texttt{score} to pass the given value.  This operation is similar to the "Change function signature" refactoring tool found in contemporary IDEs, along with other refactoring-oriented operations such as \Rename to change the name of a symbol and all references to it.  However, as shown throughout this paper, the source code algebra approach has benefits beyond refactoring.

The second modification makes the assignment of non-zero values to \texttt{bonus} only happen when indicated by \texttt{includeBonus}.  This is achieved by applying the \MakeCond operation to those statements \op{B}, with the condition expression as its argument.  Unlike \AddParam or \Rename, this kind of operation is not found in existing refactoring tools.

Going even further, Fig.~\ref{fig:intro}(c) shows that it is possible to algebraically compose these source code operations, thereby creating a new, higher-level operation.  For example, the \MakeOptional operation is defined to encapsulate the idea of ``make this code execute conditionally, based on a new parameter to the enclosing function''.  Applying it to the same original code produces the same result as applying \AddParam and \MakeCond separately.  Additionally, the algebraic properties of \MakeOptional can be derived from those of the operations it calls (\AddParam, \MakeCond, and \GetEnclosingFn).

This paper (\ref{sec:intro})~introduces source code algebra; (\ref{sec:related})~positions it against prior work; (\ref{sec:details})~sketches initial algebraic properties and their benefits; (\ref{sec:eval})~presents a feasibility probe showing accuracy improvements and one to two orders of magnitude token reduction on a non-local task; (\ref{sec:conclusions})~concludes with future directions.

%%%%%%%%%%%%%%%%%%%%%%%%%%%%%%%%%
%%
%%  RELATED WORK
%%
%%%%%%%%%%%%%%%%%%%%%%%%%%%%%%%%%

\section{Related work}
\label{sec:related}

%% Refactoring tools

Many source code algebra operations preserve program behavior, making them similar to well-studied \emph{refactoring tools}\cite{1265817, Golubev2021RefactoringSurvey}.
A central feature of source code algebra is operation composability, and while there has been some research into composable refactoring\cite{KNIESEL20049, 1553570}, it is not commonly found in actual systems.
Source code algebra can be seen as a generalization of refactoring to include changes that alter program behavior in well-controlled ways (see Section~\ref{sec:potency}).

%% Formal methods

Source code algebra is conceptually adjacent to \emph{formal methods and program verification}\cite{10.1145/27651.27653, BORBA200453, ClarkeWing1996FormalMethods, 10.1007/s00165-012-0249-0}.  The distinction is that these areas focus on the programs themselves, and their algebraic laws are used to prove properties about program construction and equivalences.  This makes them highly relevant for language designers and compiler optimizers, but not the day-to-day programming of a software engineer.  By contrast, source code algebra is focused not on provable properties of programs, but on the process of making changes to programs, i.e. editing source code.  Specifically, the programmer can only perform syntactically and semantically valid changes, each of which can be reasoned about.  This keeps the programmer "on track", similar to an applied mathematician performing a series of valid transformations to an equation.  Advanced source code algebra systems might utilise symbolic algebra\cite{vonzurgathen2013modern} to simplify or normalize operation sequences, and proof assistants\cite{Geuvers2009} to derive properties about operations.

%% Programming paradigms

\emph{Visual programming} systems\cite{Boshernitsan:CSD-04-1368,ray_vpl} have source code that is represented non-textually.  \emph{Structural and syntax-directed editors}\cite{10.1145/358746.358755, donzeau1984mentor} allow the programmer to directly modify the structure of the source code (typically the abstract syntax tree).  \emph{Intentional programming}\cite{Simonyi2006IntentionalSoftware} sought to eschew a canonical text-based syntax in favor of the AST, alongside various views/projections of the tree, and projectional editors allow editing via these views.  \emph{Language workbenches}\cite{fowler2005languageworkbenches, ERDWEG201524} are a yet-higher abstraction, which express these concepts in terms of inter-related domain-specific languages that, together with enabling tools, permit development in \emph{language-oriented programming}\cite{ward1994languageoriented, dmitriev2004languageoriented}.  By contrast to these prior approaches, source code algebra accepts text-based source code, including existing well-established programming languages, paradigms, and environments.  While structural and projectional editing ensures that source code stays structurally sound, source code algebra extends this to preserve semantic validity.  For example, the \AddParam (and related) operations mean that it's not possible for function call arguments to not match the function definition.

\emph{Aspect-oriented programming}\cite{Kiczales1997AOP} allows source code modules to be separated and composed more independently.  This helps to localize code changes and thereby reduce extensive codebase-wide modifications, but doesn't completely eliminate them, requiring the programmer to identify and make them as necessary.  By contrast, each source code algebra operation ensures that it performs all relevant modifications.

%% Patch workflows and version control systems

The \emph{stacked diffs} approach\cite{graphite-stacking, stacking-dev} to code review encourages splitting changes into smaller individual changes that, while developed cumulatively, are reviewed and landed separately.  Version control systems (VCSs) aimed at maintaining \emph{patch stacks}\cite{stacked-git, goode-bolin-sapling-2022} or with \emph{mutable histories}\cite{goode-bolin-sapling-2022, jujutsu, alden2024jujutsu} are similar, but extend this decomposition throughout the development process.  Some systems go even further, using \emph{patch algebras}\cite{roundy2005darcs, pijul-vcs, pijul-model, meunier2018pijul} to construct the codebase from the patches, rather than storing a commit graph of codebase snapshots.  These approaches are similar to source code algebra, except they remain fundamentally text-based.  For example, as explained in Section~\ref{sec:commutativity}, this means the patches in Fig.~\ref{fig:patches} will result in merge conflicts (albeit with less impact than traditional VCSs), whereas Fig.~\ref{fig:patchops} shows how source code algebra avoids this problem.

%% LLM-based code editing

Modern \emph{LLM-backed agentic coding assistants} use sophisticated indexing and retrieval\cite{engineerscodex2025cursorindexing,kinney2025cursorcontext}, plan-based context selection\cite{10.1145/3643757}, or autonomous tool-use frameworks\cite{NEURIPS2024_5a7c9475,zhang2024autocoderover} to make targeted code changes --- but in all cases the agent itself is responsible for identifying every location that must be edited and producing the necessary text edits.  By contrast, source code algebra operations automatically make all relevant changes, off-loading this burden from the LLM and allowing it to focus on determining a suitable sequence of operations.  We hypothesize that this sequence of operations is closely aligned to the higher-level conceptual plan of required code changes, and therefore well-suited for existing LLMs to generate (supported by the results in Section~\ref{sec:eval}).  These operations can then be used as input for a deterministic source code algebra system, enabling the LLM to make reliable, high-level "conceptual source code edits", rather than stochastic, low-level text edits.  Human review might also be easier when LLM output is expressed as source code algebra operations.

Coccinelle\cite{padioleau:eurosys08} addresses \emph{collateral evolution}, propagating changes via pattern-based semantic patches in a DSL (SmPL).  Comby\cite{vanTonder2019comby} generalizes this across languages, while codemod frameworks such as jscodeshift\cite{jscodeshift} and OpenRewrite\cite{openrewrite} express transformations as AST-walking scripts or composable recipes.  These share source code algebra's goal of capturing one conceptual change as a single structural unit, but each transformation is a bespoke script lacking the algebraic properties that source code algebra treats as first-class.

%%%%%%%%%%%%%%%%%%%%%%%%%%%%%%%%%
%%
%%  DETAILS
%%
%%%%%%%%%%%%%%%%%%%%%%%%%%%%%%%%%

\section{Details}
\label{sec:details}

The most basic properties of source code algebra operations are that each operation should make the \Complete set of modifications necessary for the semantic change they implement, multiple operations are \Composable (particularly for building operations at higher levels of abstraction), and operations give \Deterministic output for any given input.

To give a flavor of some further properties and their benefits, we now sketch \emph{potency} and \emph{commutativity}.  Fig.~\ref{fig:props} summarises the properties presented in this paper.

\begin{figure}
\noindent
\begin{tabularx}{\linewidth}{lX}
\toprule
\Complete      & Operations perform all necessary modifications for a semantic change \\
\Composable    & Multiple operations can be combined into a higher-level operation ($h = f \circ g$) \\
\Deterministic & For a given input, operation output is always the same \\
\midrule
\Nullipotent   & Operation does not alter behavior \\
\Pluripotent   & Operation might alter behavior \\
\midrule
\Commutative   & Operation composition is symmetric ($f \circ g \equiv g \circ f$) \\
\bottomrule
\end{tabularx}
\caption{Basic properties of source code algebra operations.}
\label{fig:props}
\end{figure}

\subsection{Potency}
\label{sec:potency}

Applying an algebraic operation to a mathematical equation results in an equation that is different, but still equivalent to the original form.  Similarly, sometimes it is the case that applying a source code algebra operation to a codebase results in code that has changed, but which remains \emph{functionally equivalent} to the original, i.e. behavior-preserving.  We say these operations are \Nullipotent.

With mathematical equations, the focus is always on nullipotent manipulations, to have confidence that the final result is a solution to the original problem.  With software engineering, this corresponds to the task of pure refactoring.

However, unlike equations, in software development it is often desirable to have functional changes, for example, new features, bug fixes, and so on.  In source code algebra, this is achieved by \Pluripotent operations, which are behavior-altering.  It is usually desirable for these behavioral changes to be tightly controlled --- each pluripotent operation should be designed to make only a single conceptual change, and be free from unintended side effects.

An advantage of source code algebra is that it brings the issue of nullipotency vs pluripotency to the fore.  Since source code is modified by applying a sequence of source code algebra operations, the potency of each operation can be considered in isolation, as well as in the context of surrounding operations.  Like equation rewriting, the decomposition facilitates reasoning about the whole, by reasoning about the parts.

The \AddParam operation in Fig.~\ref{fig:intro} is nullipotent, because the newly-added parameter is not (yet) used by the function (since it has only just been added to the scope), and as such cannot influence the behavior of the program.  The \MakeCond operation is generally pluripotent, since it causes previously always-executed statements to only sometimes be executed.  However, in this scenario it is \emph{contextually nullipotent}, because the condition is \texttt{includeBonus}, which is always \texttt{True} (by virtue of the arguments to the \AddParam operation) --- and so the affected statements continue to always be executed.

Therefore, the \MakeOptional operation is nullipotent, because it modifies the codebase to \emph{allow} the targeted statements to be conditionally executed, but without actually causing that to happen.  There would only be a change in functionality if a \emph{subsequent} operation were to modify a \texttt{score} call site to pass \texttt{False} (or add a new such call site) --- and any such subsequent operation must necessarily be pluripotent.

It would be ideal if the potency of every operation could be known, however we have not studied this question, and our intuition is that operation potency is decidable only in certain cases (and undecidable in general, due to Rice's theorem\cite{Rice1953}).  Nevertheless, even in the absence of provable potency, we still believe it to be a useful consideration, as illustrated by the above example.

\subsection{Commutativity}
\label{sec:commutativity}

A related concept is operation \Commutativity, which is when the composition $f \circ g$ of two operations $f$ and $g$ is symmetric, i.e. $f \circ g \equiv g \circ f$.  Source code operations are not always commutative, but there are benefits when they are --- which is again analogous to the situation with mathematical function composition.

For example, consider the pair of text patches shown in Fig.~\ref{fig:patches}(a) and (b); the first changes the name of a function parameter, while the second changes the order of the parameters.  With traditional text-based patch or revision control systems, attempting to apply both of these patches leads to a \emph{merge conflict}, requiring manual intervention to resolve the desired final state shown in Fig.~\ref{fig:patches}(c).

By contrast, consider if these changes are instead represented by (or alternatively, were caused by) the operations shown in Fig.~\ref{fig:patchops}, rather than by text-based patches.  Here, changing the parameter name is achieved by the $\Rename(\op{A}, \mathtt{message})$ operation, while the parameter order is changed by $\MoveParam(\op{A}, +1)$ (to move the parameter's position to be 1 later in the parameter list).

Fig.~\ref{fig:patchops} illustrates that these operations are commutative, and can be applied in either order to achieve the desired output without any merge conflicts.  Although a simple example, this concept can be extended to other situations where the order of changes is significant, for example rebasing or merging VCS branches, or linearized commit queues.  This shows how commutative source code algebra operations can provide benefits that are unavailable or difficult in traditional text-based systems.

%%%%%%%%%%%%%%%%%%%%%%%%%%%%%%%%%
%%
%%  EVALUATION
%%
%%%%%%%%%%%%%%%%%%%%%%%%%%%%%%%%%

\section{Evaluation}
\label{sec:eval}

We have developed an open-source proof-of-concept source code algebra implementation, known as \emph{Source Code Algebraic System (SCAS)}\cite{scas}.  Source code syntax is parsed with tree-sitter\cite{https://doi.org/10.5281/zenodo.4619183}, ``enriched'' with semantic information derived from (usually language-specific) static analysis, and the resulting tree stored as a MongoDB\cite{mongodb-manual-8} collection.  It exposes operations as composable Unix-style programs, and integrates as a tool an LLM agent can call.

SCAS relies on the AST enrichment, which currently uses Spoon\cite{pawlak:hal-01169705} for Java, but is moving to support more languages via \texttt{semanticToken}-enabled LSP\cite{lsp2026} servers.

A feasibility probe was conducted to investigate if SCAS can be used as-is by existing agentic LLM frameworks, requiring only access to the tool and its associated documentation (i.e. via in-context learning, without model fine-tuning or special model/agent architectures).

\begingroup

\newsavebox{\boxpatcha}
\begin{lrbox}{\boxpatcha}
\begin{mycode}%
\rem{-\kw{void} fn(\kw{char*} \REM{msg}, \kw{int} type);}\\
\add{+\kw{void} fn(\kw{char*} \ADD{message}, \kw{int} type);}
\end{mycode}%
\end{lrbox}

\newsavebox{\boxpatchb}
\begin{lrbox}{\boxpatchb}
\begin{mycode}%
\rem{-\kw{void} fn(\REM{\kw{char*} msg, ~}\kw{int} type);}\\
\add{+\kw{void} fn(\kw{int} type\ADD{, \kw{char*} msg});}
\end{mycode}%
\end{lrbox}

\newsavebox{\boxpatchc}
\begin{lrbox}{\boxpatchc}
\begin{mycode}%
\rem{-\kw{void} fn(\REM{\kw{char*} msg, ~}\kw{int} type);}\\
\add{+\kw{void} fn(\kw{int} type\ADD{, \kw{char*} message});}
\end{mycode}%
\end{lrbox}

\begin{figure}
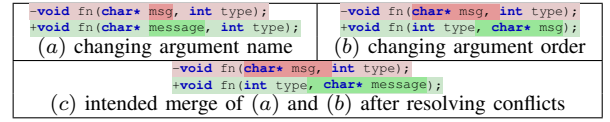

\centering
{
\footnotesize
\begin{tabular}{|c|c|}%
\hline
\usebox{\boxpatcha} & \usebox{\boxpatchb} \\
{$(a)$ changing argument name} & {$(b)$ changing argument order} \\
\hline
\multicolumn{2}{|c|}{\usebox{\boxpatchc}} \\
\multicolumn{2}{|c|}{{$(c)$ intended merge of $(a)$ and $(b)$ after resolving conflicts}} \\
\hline
\end{tabular}%
}
\caption{Merging text-based patches can result in conflicts that require manual resolution.}
\label{fig:patches}
\end{figure}

\newsavebox{\boxpatchopa}
\begin{lrbox}{\boxpatchopa}
\begin{mycode}%
\kw{void} fn(\kw{char*} \tgt{A}{msg}, \kw{int} type);
\end{mycode}%
\end{lrbox}

\newsavebox{\boxpatchopaa}
\begin{lrbox}{\boxpatchopaa}
\begin{mycode}%
\kw{void} fn(\kw{char*} \tgt{A}{msg}, \kw{int} type);
\end{mycode}%
\end{lrbox}

\newsavebox{\boxpatchopab}
\begin{lrbox}{\boxpatchopab}
\begin{mycode}%
\kw{void} fn(\kw{char*} \tgt{A}{\REMST{msg}\ADD{message}}, \kw{int} type);
\end{mycode}%
\end{lrbox}

\newsavebox{\boxpatchopac}
\begin{lrbox}{\boxpatchopac}
\begin{mycode}%
\kw{void} fn(\REMST{, }\kw{int} type\ADD{, ~}\mov{\kw{char*} message});
\end{mycode}%
\end{lrbox}

\newsavebox{\boxpatchopba}
\begin{lrbox}{\boxpatchopba}
\begin{mycode}%
\kw{void} fn(\kw{char*} \tgt{A}{msg}, \kw{int} type);
\end{mycode}%
\end{lrbox}

\newsavebox{\boxpatchopbb}
\begin{lrbox}{\boxpatchopbb}
\begin{mycode}%
\kw{void} fn(\REMST{, }\kw{int} type\ADD{, ~}\mov{\kw{char*} \tgt{A}{msg}});
\end{mycode}%
\end{lrbox}

\newsavebox{\boxpatchopbc}
\begin{lrbox}{\boxpatchopbc}
\begin{mycode}%
\kw{void} fn(\kw{int} type, \kw{char*} \REMST{msg}\ADD{message});
\end{mycode}%
\end{lrbox}

\newsavebox{\boxpatchopc}
\begin{lrbox}{\boxpatchopc}
\begin{mycode}%
\kw{void} fn(\kw{int} type, \kw{char*} message);
\end{mycode}%
\end{lrbox}

\begin{figure}
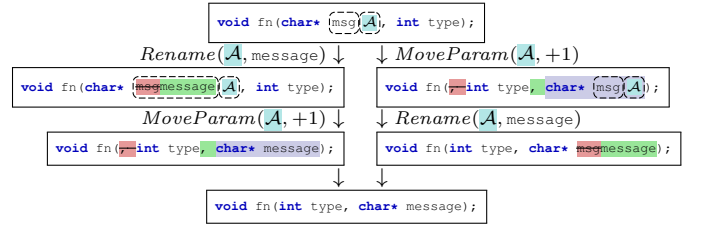

\centering
{
\footnotesize
\begin{tabular}{@{}rl@{}}%

\multicolumn{2}{c}{\fbox{\usebox{\boxpatchopa}}} \\

\scasdownarrowafterinner{\Rename(\op{A}, \scastt{message})} &
\scasdownarrowinner{\MoveParam(\op{A}, +1)} \\

\fbox{\usebox{\boxpatchopab}} & \fbox{\usebox{\boxpatchopbb}} \\

\scasdownarrowafterinner{\MoveParam(\op{A}, +1)} &
\scasdownarrowinner{\Rename(\op{A}, \scastt{message})} \\

\fbox{\usebox{\boxpatchopac}} & \fbox{\usebox{\boxpatchopbc}} \\

{$\downarrow$} & {$\downarrow$} \\

\multicolumn{2}{c}{\fbox{\usebox{\boxpatchopc}}} \\

\end{tabular}%
}
\caption{Commutative operations can be applied in any order without introducing conflicts.}
\label{fig:patchops}
\end{figure}

\endgroup

The experimental setup consisted of various LLMs in a ReAct framework\cite{yao2023reactsynergizingreasoningacting}, implemented using LangChain\cite{langchain}, with tool access to SCAS (including a built-in contextual \texttt{help} operation).  The task is a single cross-file symbol rename (a Java class method), in a synthetic codebase\footnote{Generated with the assistance of the Cursor AI coding tool.} of $\sim$5.5k lines of code ($\sim$500kb, $\sim$200k tokens, across 8 files), with deliberately ambiguous symbol names (such that naive text-based replacement will fail).

This setup was chosen as a minimum viable demonstration for the feasibility of an LLM in-context learning the concept of source code algebra, and then subsequently applying appropriate SCAS operations to modify the codebase algebraically, without text edits.  A familiar task (symbol rename) is used to avoid problems where the LLM misunderstands the request.

The system is compared against two baselines: (1) the "plain" LLM with one-shot prompting (i.e. without an agentic framework, tools, or other strategies such as chain of thought prompting), and reduced input codebase size of $\sim$20k tokens (due to context window constraints), and (2) the Cursor\cite{cursor-ai} coding assistant (v2.5.20) running in Agent mode.

The SCAS implementation, tool definitions, prompts, evaluation codebase, and full input/output transcripts are available in a branch of the SCAS repository\cite{scas}.

\subsection{Results}

% Deliberately outside the group, so that the main text can use them.
\newcommand{\yes}{{\color{green}\ding{51}}\xspace}
\newcommand{\no}{{\color{red}\ding{55}}\xspace}
\newcommand{\fixup}{{$\sim$}\xspace}

\begingroup

\begin{figure*}[]
\centering
\begin{tabular}{
  l
    !{\hspace{0.5em}}
  c@{}c@{}c@{\hspace{1em}}r@{ }r
    !{\hspace{0em}}
  c@{}c@{}c@{\hspace{1em}}r@{ }r
    !{\hspace{1em}}
  c@{}c@{}c@{\hspace{1em}}r@{ }r
    !{\hspace{1em}}
  c@{}c@{}cr@{ }r
}
\toprule

&
\multicolumn{5}{c}{\textbf{One-shot (full)}} &
\multicolumn{5}{c}{\textbf{One-shot (small${}^*$)}} &
\multicolumn{5}{c}{\textbf{Cursor (full)}} &
\multicolumn{5}{c}{\textbf{SCAS (full)}}
\\

\textbf{Model} &
\multicolumn{3}{@{}c@{}}{\textbf{Result}} & \multicolumn{2}{@{}c@{}}{\textbf{Tokens}} &
\multicolumn{3}{@{}c@{}}{\textbf{Result}} & \multicolumn{2}{@{}c@{}}{\textbf{Tokens}} &
\multicolumn{3}{@{}c@{}}{\textbf{Result}} & \multicolumn{2}{@{}c@{}}{\textbf{Tokens}} &
\multicolumn{3}{@{}c@{}}{\textbf{Result}} & \multicolumn{2}{@{}c@{}}{\textbf{Tokens}}
\\

\midrule

gpt-4o-mini &
\multicolumn{3}{c}{\textit{n/a}} & \multicolumn{2}{c}{\textit{n/a}} &
\no & \no & \no & 25k & (123\%) &
\multicolumn{3}{c}{\textit{n/a}} & \multicolumn{2}{c}{\textit{n/a}} &
\yes & \no & \no & 11k & (5.48\%)
\\

gpt-5-mini &
\multicolumn{3}{c}{\textit{n/a}} & \multicolumn{2}{c}{\textit{n/a}} &
\yes & \yes & \yes & 51k & (255\%) &
\yes & \no & \no & 545k & (272\%) &
\yes & \yes & \yes & 19k & (9.66\%)
\\

claude-4.5-sonnet &
\multicolumn{3}{c}{\textit{n/a}} & \multicolumn{2}{c}{\textit{n/a}} &
\yes & \yes & \yes & 60k & (302\%) &
\yes & \yes & \yes & 1,124k & (562\%) &
\yes & \yes & \yes & 23k & (11.3\%)
\\

claude-4.5-haiku &
\multicolumn{3}{c}{\textit{n/a}} & \multicolumn{2}{c}{\textit{n/a}} &
\no & \no & \no & 30k & (151\%) &
\yes & \no & \no & 434k & (217\%) &
\yes & \yes & \yes & 19k & (9.30\%)
\\

claude-4.5-opus &
\multicolumn{3}{c}{\textit{n/a}} & \multicolumn{2}{c}{\textit{n/a}} &
\yes & \yes & \yes & 60k & (302\%) &
\no & \no & \no & 109k & (54\%) &
\yes & \yes & \yes & 18k & (9.22\%)
\\

gpt-5.2 &
\fixup & \fixup & \no & 204k & (102\%) &
\yes & \yes & \no & 49k & (246\%) &
\fixup & \fixup & \fixup & 778k & (389\%) &
\yes & \yes & \yes & 7k & (3.71\%)
\\

\bottomrule
\end{tabular}

\caption{%
Experimental results from various LLM-based systems to solve a synthetic coding exercise.  The full input source code is $\sim$200k tokens, while the small${}^*$ source code is reduced by $\sim$90\% to $\sim$20k tokens.  \textbf{One-shot} is a plain LLM without tools, agentic framework, or any other strategies.  \textbf{Cursor} is the Cursor coding assistant in Agent mode.  \textbf{SCAS} is the POC source code algebra implementation as a ReAct style tool in LangChain, including in-context learning of how to use the SCAS system.  Each model-system combination was run 3 times.  The output was inspected and a \textbf{Result} of \yes~indicates that the source code was correctly modified, \fixup~indicates mostly correct output (requiring only minor corrections), and \no~indicates that the source code was not correctly modified.  \textbf{Tokens} indicates the total input and output tokens used by the system on the median run, including as a percentage of the input codebase size (200k for full, or 20k for small${}^*$).  For One-shot (full), \textit{n/a} indicates that the model's context window is too small for the input codebase.  For Cursor, \textit{n/a} indicates that the model is not available in the system.  The source code algebra system compares favorably to the text-editing baselines, producing correct results more often and with fewer tokens.
}
\label{fig:results}
\end{figure*}

\endgroup

Fig.~\ref{fig:results} shows the results of the experiment.  Each model-system pair was executed 3 times.  Output code was inspected by the author and judged for correctness according to:

\begin{itemize}
\item \yes   $\implies$ the source code was correctly modified.
\item \fixup $\implies$ the resulting source code was mostly correct, requiring only minor corrections, e.g. the model may have made extraneous (albeit correct) changes, or may have elided some unchanged sections of code.
\item \no    $\implies$ the source code was not correctly modified.
\end{itemize}

\noindent The total number of tokens (input and output) were recorded; the minimum and maximum of the 3 runs were discarded, leaving the median token usage.

Unsurprisingly, the One-shot performance is poor.  While it can usually produce a correctly renamed method definition, it fails to update the corresponding call sites in all but the most capable model tested (gpt-5.2).  Even when this can be achieved, the tokens required grows in proportion to the input source code (with a coefficient of $\sim$2, since the complete source code must be input and output).  Cursor is sometimes able to correctly update all locations of the source code, however, doing so requires a large number of tokens --- up to $\sim$5 times the size of the input source code.

When using SCAS the agent successfully completes the task, except when using the least-capable model (gpt-4o-mini).  It also requires strikingly fewer tokens with SCAS: around 7k~--~23k tokens, or $\sim$3\%~--~11\% of the codebase size, which is roughly one to two orders of magnitude lower than Cursor's $\sim$217\%~--~562\%.  SCAS requires fewer tokens because with it, the model doesn't need to identify and make all the cascading changes --- whereas Cursor likely does require this.

Additionally, the SCAS token usage remains low despite also including the agent's exploratory in-context learning of how to use the SCAS system, suggesting that these learning costs are low.

The SCAS token usage is expected to scale with the size of the immediately relevant section of source code --- rather than the size of all modified code, or the whole codebase --- because SCAS offloads the requirement for the model or agent to find and update all necessary code sites.

These preliminary results are consistent with our hypothesis that source code algebra is a useful approach for modifying source code, and the hypothesis that LLMs are able to translate their high-level plan into source code algebra operations.  This is encouraging evidence that the approach warrants further research, especially as a substrate for coding agents.

\emph{Limitations:} This is a feasibility probe --- a single non-local task, on one synthetic codebase, in one language, with limited repetition and inspection-judged correctness.  Broader evaluation across benchmarks, languages, and baselines is left to future work.

%%%%%%%%%%%%%%%%%%%%%%%%%%%%%%%%%
%%
%%  CONCLUSIONS AND FUTURE WORK
%%
%%%%%%%%%%%%%%%%%%%%%%%%%%%%%%%%%

\section{Conclusions and future work}
\label{sec:conclusions}

This paper has proposed \emph{source code algebra}: a way of modifying a codebase by applying a sequence of logical algebraic operations to it, rather than by editing text.  The approach treats code modifications as composable units with reasoning-friendly properties --- including nullipotency and commutativity --- analogous to how mathematicians rewrite equations using algebraic identities.

An initial probe with our proof-of-concept implementation (SCAS) shows that LLM agents are able to drive source code algebra to complete a non-local, cross-file code change with high accuracy, and roughly one to two orders of magnitude fewer tokens than text-based baselines.  While preliminary, this is consistent with the hypothesis that having LLM agents emit algebraic operations, rather than rewriting or text editing source code, is a promising direction for LLM coding agents.

Several directions warrant further investigation.  The operator library is currently small and Java-focused; expanding both its breadth and language coverage is a prerequisite for detailed evaluation on standard benchmarks such as SWE-bench\cite{jimenez2024swebench}.  The properties sketched in Section~\ref{sec:details} warrant formal study --- particularly the decidability of nullipotency and commutativity, conditions for their automated deduction, identification of other potentially-useful properties (or a full characterization of the formal algebra), and determination of a minimal sufficient set of operations.  Source code algebra is equally applicable to human programmers, especially if supported by interactive tooling, with potential benefits for code authoring, code review, and disciplined program evolution that remain to be investigated empirically.

We believe source code algebra is a research direction worth pursuing, both as a substrate for agentic code editing systems and as a way of bringing some of mathematics' methodical discipline to the way software is changed.

%%%%%%%%%%%%%%%%%%%%%%%%%%%%%%%%%
%%
%%  REFERENCES
%%
%%%%%%%%%%%%%%%%%%%%%%%%%%%%%%%%%

\newpage

% trigger a \newpage just before the given reference
% number - used to balance the columns on the last page
% adjust value as needed - may need to be readjusted if
% the document is modified later
%\IEEEtriggeratref{8}
% The "triggered" command can be changed if desired:
%\IEEEtriggercmd{\enlargethispage{-5in}}

% can use a bibliography generated by BibTeX as a .bbl file
% BibTeX documentation can be easily obtained at:
% http://mirror.ctan.org/biblio/bibtex/contrib/doc/
% The IEEEtran BibTeX style support page is at:
% http://www.michaelshell.org/tex/ieeetran/bibtex/
\bibliographystyle{IEEEtran}
% argument is your BibTeX string definitions and bibliography database(s)
\bibliography{IEEEabrv,scas}

% that's all folks
\end{document}